\documentstyle[twocolumn,seceq]{jpsj}
\renewcommand\figureheight[1]{\vspace{24pt}\mbox{\rule{0cm}{#1}}}

\def\runtitle{Direct observation of the spin gap excitations in
SrCu$_2$(BO$_3$)$_2$}
\def\runauthor{Hiroyuki {\sc Nojiri}, Hiroshi {\sc Kageyama},
Kenzo {\sc Oniduka}, Yutaka {\sc Ueda}, Mitsuhiro {\sc
Motokawa}}

\title
{
Direct observation of the multiple spin gap excitations in
two-dimensional dimer system SrCu$_2$(BO$_3$)$_2$}

\author
{Hiroyuki {\sc Nojiri}, Hiroshi {\sc Kageyama}$^1$, Kenzo {\sc
Onizuka}$^1$, \\Yutaka {\sc Ueda}$^1$ and Mitsuhiro {\sc
Motokawa}
}

\inst
{Institute for Materials Research, Tohoku University, Katahira
2-1-1, Sendai 980-8577, Japan

$^1$Institute for Solid State Physics,
University of Tokyo, Roppongi 7-22-1, Tokyo
106-8666, Japan
}

\recdate
{
\
}

\abst
{
Various spin gap excitations have been observed in the
two-dimensional dimer system SrCu$_2$(BO$_3$)$_2$ by means of
submillimeter wave ESR. The zero-field energy gap of the lowest
spin gap excitation shows a splitting into two triplet modes and
the energy splitting clearly depends on the magnetic field
orientation when a field is rotated in the {\mib {ac}}-plane. A
zero-field splitting is also found between the {\it S}\(_{z}\)=+1
and {\it S}\(_{z}\)=-1 branches of each triplet. These behaviors
are qualitatively explained by considering the anisotropic
exchange coupling of inter-dimer and intra-dimer, respectively.
The averaged value of the lowest spin gap energy is determined
to be 722$\pm$2 GHz(34.7 K). We have also found
the second spin gap excitation at 1140 GHz(54.7 K), which
indicates that the inter-dimer coupling is
significantly strong. Besides these modes, a number of gapped
ESR absorption are found and we propose that these multiple
magnetic excitations are caused by the localized nature of the
excited state in the present system. }

\kword
{
SrCu$_2$(BO$_3$)$_2$, spin gap, ESR, magnetic excitation, high
magnetic field }

\begin{document}
\sloppy
\maketitle

Low-dimensional quantum spin systems with a finite spin gap have been a subject of intense and advancing research in this
decade. One of the factors lies in prominent progress in solid state chemistry enabling a tailoring of various low-dimensional spin
systems for the experimental verification of a concept brought by a theoretical study. 
Recently, a strontium copper boron oxide SrCu$_2$(BO$_3$)$_2$ with a layered structure has been added to the list of
two-dimensional(2D) system with a spin-gapped ground state.\cite{rf:1}  A prominent characteristic is that this material can be
regarded as a model system with an exactly known ground state. SrCu$_2$(BO$_3$)$_2$ has a tetragonal unit cell and all Cu$^{2+}$ ions
that carry spin{\it S}=1/2 are located at equivalent sites.\cite{rf:2}  In a CuBO$_3$ layer sandwiched by layers of Sr ions, a dimer
unit is made up of the neighboring pair of planer rectangular CuO$_4$ and these dimers connect orthogonally by way of a triangular
planer BO$_3$, providing a unique 2D network of {\mib S}=1/2 as shown in the inset of Fig. 1. This 2D lattice is topologically
equivalent to a 2D square lattice with additional alternating diagonal interactions, for which the direct product of the singlet pairs
is the exact eigenstate as proven by Shastry and Sutherland.\cite{rf:3} This model for SrCu$_2$(BO$_3$)$_2$ thus can be considered as a
2D analogue of the Majumdar-Ghosh model for a one-dimensional zigzag spin chain with some stringent condition.\cite{rf:4} 
The role of the next-nearest-neighbor
interaction in the {\mib {ac}}-plane {\mib J}$_2$, which brings a frustration into the system, has been also studied
theoretically by Miyahara and Ueda.\cite{rf:5,rf:6} They estimated the intra-dimer coupling of {\mib J}=100 K and the
next-nearest-neighbor one of {\mib J}$_2$=68 K using the experimentally obtained Weiss temperature of -92.5 K and the energy
gap of 30 K. It should be noted that the ratio {\mib J}$_2$/{\mib {J}}=0.68 is just below the critical value of {\mib J}$_2$/{\mib
J}=0.70 between the spin gap state and the N${\acute e}$el-ordered state. Thus this material resides near the border between these two
states.\cite{rf:6,rf:14}

 Another interesting feature of this material appears in the
magnetization curve with quantized plateaux at one-quarter, one-eighth\cite{rf:1},and possibly one-tenth\cite{rf:7} of the saturated
moment. These magnetic plateaux have been realized in terms of the extremely localized nature of the low-lying excited
triplets.\cite{rf:6}  It means that excited triplets cannot propagate freely in the 2D lattice. As a
result, it is considered that these triplets organize a regular lattice made up of spatially separated triplets. 

Hence, it will be very interesting to study the field dependence of the magnetic excitation of this system.  However, since the
critical field is above 20 T, only a few methods can be applied to investigate of the magnetic excitation in such high fields. Among
many different methods, ESR has been established its unique status as a probe of a magnetic excitation especially to study its field
variation.\cite{rf:8,rf:9,rf:10} Thus we have performed a systematic study of magnetic excitations of SrCu$_2$(BO$_3$)$_2$ by
using a submillimeter wave electron spin resonance. The present work is the first paper of our successive articles and the main issue
is to investigate the magnetic excitations from the exact dimer ground state of the system. In the following, ESR results will be shown
after the brief mention for the experimental procedure and finally the energy diagram and localized nature of magnetic excitation is
discussed.

 Submillimeter wave ESR measurements have been performed up to 1 THz and in
pulsed magnetic fields up to 30 T. A far-infrared laser, backward travelling wave tubes and Gunn oscillators have been employed as the
radiation source. An InSb is used as a detector. High-purity bulk single crystals of SrCu$_2$(BO$_3$)$_2$ were grown by the travelling
solvent floating zone (TSFZ) method\cite{rf:11} By means of Laue
X-ray back-reflection, the grown materials was checked and the crystal axes were determined. 

ESR spectra at 1.6 K are
depicted in Fig. 1 for {\mib H}${\parallel}$a. Among many different kinds
of ESR signals, a set of two resonance peaks marked by arrows shows characteristic frequency dependence. Namely, as the frequency is
increased, the resonance field first decreases and then increases again at higher frequency. The resonance fields measured up to 1 THz
are summarized as the frequency field diagram in figures 2(a) and 2(b) for {\mib H}${\parallel}$c and {\mib H}${\parallel}$a,
respectively. As shown in these figures, the above mentioned ESR signals marked by closed circles(T1$_u$ and T1$_d$) and closed
rectangles(T2$_u$ and T2$_d$) are associated with a zero-field energy gap of about 720 GHz. The value of the energy gap is close to the
gap estimated so far by different methods.\cite{rf:1,rf:7,rf:12,rf:13} These facts indicate clearly that the two modes correspond to
the transitions between the ground singlet state and the first excited triplet state(we call these modes singlet-triplet transition
hereafter). As shown in Figs. 2(a) and 2(b), we have observed two branches of {\it S}\(_{z}\)=+1 and {\it S}\(_{z}\)=-1 for each
triplet mode T1 or T2(we use the notation such as T1 when we treat the T1$_u$ and T1$_d$ as a set of triplet). The {\it S}\(_{z}\)=0 branch cannot be detected by our
field scanning ESR. 

These lowest spin gap excitations T1 and T2 shows a complicated structure. For {\mib H}${\parallel}$a, a
zero-field splitting exists between two branches of {\it S}\(_{z}\)=+1 and {\it S}\(_{z}\)=-1. The modes T1$_u$ and T2$_u$ or T1$_d$
and T2$_d$ are degenerated at zero-field and the center of {\it S}\(_{z}\) =+1 and {\it S}\(_{z}\)=-1 branches are identical between T1
and T2 modes. A small splitting observed at high fields between T1$_u$ and T2$_u$ or between T1$_d$ and T2$_d$ cannot be explained by
the anisotropy of {\mib g}-value in the {\mib c}-plane because all Cu sites are equivalent with respect to magnetic fields for {\mib
H}${\parallel}$a in the known crystal structure at room temperature.\cite{rf:2} Thus we infer that a small lattice distortion occurs at
low temperature, however a detailed structure analysis at low temperature is required to clarify this point.  On the other hand, for
{\mib H}${\parallel}$c, no zero-field splitting exists between two branches of {\it S}\(_{z}\)=+1 and {\it S}\(_{z}\)=-1, while two
triplets T1 and T2 show a splitting at zero-field. These characteristic behaviors will be explained later by considering the
anisotropic exchange coupling of inter-dimer and intra-dimer. The slope of these modes for each field direction is nearly identical
with that of the paramagnetic resonance signals indicated by the dashed lines in Figs. 2(a) and 2(b). The {\mib g}-value are
determined as {\mib g}$_a$=2.05 and {\mib g}$_c$=2.28 by the paramagnetic resonance peak positions at 25 K. From Figs. 2(a) and
2(b), we determine the average value of the lowest spin gap as 722$\pm$22 GHz(34.7 K).

 This assignment of T1 and T2 to the singlet-triplet transition is
further confirmed by the temperature dependence of ESR spectra as shown in Fig. 3. The intensity of two peaks marked by
the arrow decreases as the temperature is increased. This behavior indicates that these signals are the transitions from the ground
state of the system. As is well known, the transition between the ground singlet state and the excited triplet state is forbidden in
principle as the magnetic dipole transition observed in ESR. However, the presence of a non-secular term such as
Dzyaloshinsky-Moriya(DM) interaction, non-equivalent {\mib g}-tensors or anisotropic exchange interaction makes it possible to observe
this transition by means of ESR. We speculate that the anisotropic
exchange is the origin of the breaking down of the selection rule in the present system and the reason is given in the following.  

One of the evidence of the anisotropic intra-dimer exchange interaction is
the zero-field splitting between the two branches of {\it S}\(_{z}\)=+1 and {\it S}\(_{z}\)=-1 for each triplet mode(for example
between T1$_u$ and T1$_d$ or between T2$_u$ and T2$_d$) for {\mib H}${\parallel}$a as shown in Fig. 2(b). Let us consider an isolated
dimer of {\it S}=1/2 coupled by an antiferromagnetic exchange interaction with exchange anisotropy. Since each Cu is bridged by a
planer CuO$_4$ unit which lies parallel in the {\mib c}-plane, we can expect that {\mib {J}}$_{cc}$ deviates from {\mib {J}}$_{aa}$ in
the present system, where {\mib {J}}$_{cc}$ and {\mib {J}}$_{aa}$ are the components of exchange coupling along the {\mib c}-axis and
{\mib a}-axis, respectively. By diagonalizing the Hamiltonian of a dimer with such anisotropic exchange coupling, we can show that the
splitting at a zero-field exists between {\it S}\(_{z}\)=+1 and {\it S}\(_{z}\)=-1 branches for {\mib H}${\perp}$c and that no
splitting exists for {\mib H}${\parallel}$c between these two branches. This field orientation dependence is consistent with the
experimental results of the splitting between {\it S}\(_{z}\)=1 and {\it S}\(_{z}\)=-1 branches shown in Figs. 2(a) and 2(b).  As is
well known, the order of the anisotropic exchange AE can be estimated roughly by the relation AE$\sim $($\Delta ${\mib g}/{\mib
g})$^2${\mib J}$_i$, where $\Delta
${\mib g}/{\mib g} is the anisotropy of the {\mib g}-value normalized by the average of {\mib g} and {\mib J}$_i$ is the isotropic
part of exchange constant. By using  {\mib g}$_a$=2.05, {\mib g}$_c$=2.28 and {\mib {J}}$_i$${\approx }${\mib {J}}=100 K, we evaluate
AE$\sim $1 K. This value is  comparable with the zero-field splitting of about 30 GHz in Fig. 2(b) and thus we consider that the
anisotropic exchange is the origin of the above mentioned splitting and this term makes the singlet-triplet transition allowable for
ESR.

 Next let us discuss the parallel
splitting of the lowest energy spin gap excitation for {\mib H}${\parallel}$c into T1 and T2 modes. The splitting between T1 and T2 can
not be attributed to the difference of {\mib g}-value because the splitting exists even at zero-fields for {\mib H}${\parallel}$c and
both modes are parallel to each other as shown in Fig. 2(a). Such splitting can be expected when an inter-dimer interaction {\mib
J}$^{\prime}$ exists. Since the splitting depends on the field orientation, we speculate that the splitting is caused by
the anisotropy of {\mib J}$^{\prime}$ and it is not related to the absolute total value of {\mib
J}$^{\prime}$. To confirm this point, we have measured angular dependence of two peaks T1$_d$ and T2$_d$ at 428.9 GHz as shown in
Fig. 4. By using the angular dependence of {\mib g}-value, the zero-field energy of two triplet is evaluated as a function of field
orientation as shown in the inset of Fig. 4. It should be noted
that the splitting shows a sinusoidal angular dependence. Thus we confirmed that the splitting between T1 and T2 for {\mib
H}${\parallel}$c is caused by the anisotropic inter-dimer exchange interaction.

 When a finite {\mib J}$^{\prime}$ exists, we can
expect a second excited state which shows a splitting of about {\mib J}$^{\prime}$ from the lowest triplet state. In practice, the
second lowest singlet-triplet transition is found and the energy gap is about 1140 GHz(see closed triangles in Figs. 2(a) and
2(b)). This second triplet state can be understood by using a simple model as follows. When {\mib
J}$^{\prime}${\mib {<J}}, we can treat a two dimer coupled by {\mib J}$^{\prime}$ as a dimer of two {\it S}=1 spins.  In this case,
the ground state({\it E}=0), the first excited state({\it E}={\mib J}$^{\prime}$) and the second excited state({\it {\it E}}=3{\mib
J}$^{\prime}$) are {\it S}=0 singlet, {\it S}=1 triplet and {\it S}=2 quintet, respectively, where {\it E} is the energy of each
level. Thus we can expect that the second triplet state is located at the {\it E}={\mib J}$^{\prime}$ above the lowest triplet state.
The experimentally observed splitting of 418 GHz=1140 GHz -722 GHz is the measure of inter-dimer coupling. If we neglect the
third-neighbor interaction in the {\mib c}-plane, {\mib J}$^{\prime}$ is roughly the half of {\mib J}$_2$=68 K(1416 GHz).\cite{rf:6}
The agreement between these two estimations of {\mib J}$^{\prime}$ are very poor because our model is too simple. However, these
results show that inter-dimer interaction is considerably large in the present material. This second energy gap was also observed by
the neutron scattering experiments which was made very recently.\cite{rf:15} Thus we conclude that the splitting between T1 and T2 for
{\mib H}${\parallel}$c as well as the existence of the second spin gap for both {\mib H}${\parallel}$c and {\mib H}${\parallel}$a are
attributable to the large {\mib J}$^{\prime}$ in the present system. 
 
When we look at the frequency field diagram shown in Figs. 2(a) and 2(b),
we notice that a number of ESR signals are observed besides the lowest and the second lowest energy modes of the singlet-triplet
transition. We categorize these signals into following two types; (1) strong ESR absorption appearing just below the critical
field(closed diamond), and (2)other weak ESR signals marked by open rectangles. To investigate the origin of the strong ESR mode (1),
we measured the temperature dependence of ESR spectra as shown in Fig. 3. The signal located at 22 T is a paramagnetic resonance.
Since the intensity of all ESR peaks observed at 1.6 K decreases as the temperature is increased, these signals are considered as the
excitations from the ground state. It should be also noted that the intensity of the peaks appearing around 15-17 T is
much stronger than that of the singlet-triplet transition and is comparable with that of paramagnetic signal. Moreover, these signals
are not related to the magnetic excitations at a plateau because the fields are well below the critical field {\mib H}$_c$, where the
lowest triplet state crosses with the singlet ground state. These facts indicate that the strong signals are related to the ground
state but are not a forbidden transition such as singlet-triplet transition. Hence, we infer that the ground state of the system near
the critical field is not a perfect nonmagnetic singlet state; i.e. a part of the singlet is broken and as a consequence a magnetic
ground state is recovered locally in some part of the system for the finite mixing between the ground singlet state and the excited
states. In practice, the bending of triplet mode T1$_d$ and T2$_d$ are observed above 15 T as shown in Fig. 2(a) and the magnetization
is non-zero even at 0.5 K in the field well below {\mib H}$_c$.\cite{rf:7} 

 Finally, we discuss the origin of the weak signals marked by open rectangles which are categorized as (2). It seems that many of
them lie almost parallel to the singlet-triplet transition T1 or T2. It means that these ESR signals have finite zero-field energy gap
and thus they are the transitions between the ground state and the excited state. We speculate that these weak absorption are the
localized multiple spin gap excitation caused by the extremely localized nature of the excited state proposed
theoretically.\cite{rf:6} Usually, an excited triplet in the dimerlized spin-gap system can propagate for the finite inter-dimer
coupling and thus a significant dispersion is observed. In the present system, a triplet excitation is extremely localized and its
propagation is very limited. In this case, we expect that a dispersive mode may be replaced by a number of excitations, which consists
of nearly degenerated discrete energy levels. Namely, we can expect a state that triplets are coupled as dimer, trimer, quadramer etc.
In this case, the excitation energy of each level may be given by a sum of the formation energy of the isolated  triplet, the
interactions with other triplets located nearby and the Zeeman energy. This idea can explain why the most of weak ESR peaks appear
above the {\it S}\(_{z}\)=-1 branch of the second lowest singlet-triplet transition(closed triangles). It also explains that the
number of the peaks increases when a field approach to the critical field {\mib H}$_c$. It is because the number of the surrounding
triplets are increased around the {\mib H}$_c$. As is well known, such localized excitations with discrete energy levels have been
known as a spin cluster excitation for Ising system, however, it is found for the first time for Heisenberg system.\cite{rf:16} A
further theoretical and experimental investigation is desirable to clarify our proposal. 

To summarize, we have directly found the singlet-triplet transition of the SrCu$_2$(BO$_3$)$_2$ by submillimeter wave ESR. Two sets
of well-defined triplet modes are found and the energy gaps are evaluated to be 722$\pm$2 GHz(34.7 K) and 1140 GHz(54.7 K). In
additions to these transitions, we have also observed multiple spin gap excitations, which may be realized by the extremely localized
nature of the excited state.

\section*{Acknowledgements}
	This work was partly supported by Grant-in-Aid of Ministry of Education, Science, Sports and Culture.

Figure captions

\begin{figure}
\figureheight{0cm}
\caption{Examples of ESR spectra at 1.6 K for {\mib H}${\parallel}$a. Arrows indicate two lowest
triplets T1 and T2. The suffix u and d indicate the {\it S}\(_{z}\)=+1 and {\it S}\(_{z}\)=-1 branches, respectively.  Closed triangle
and open rectangles are the second lowest triplet and other multiple weak signals, respectively. The inset shows the
schematic crystal structure in the  {\mib c}-plane. The dotted lines shows the unit cell. }
\label{fig:1}
\end{figure}

\begin{figure}
\figureheight{0cm}
\caption{The frequency-field diagrams at 1.6 K for {\mib H}${\parallel}$c (a) and for {\mib H}${\parallel}$a (b). The dashed
line indicates the position of paramagnetic resonance. Closed circles and closed rectangles are the two triplet modes T1 and T2.
 Closed triangle is the second lowest triplet.  Closed diamond and open rectangles are other strong and weak signals, respectively
as mentioned in the text. The magnetization curve of single crystal measured at 0.5 K taken from
Ref. 8 is also plotted for {\mib H}${\parallel}$a. }
\label{fig:2}
\end{figure}

\begin{figure}
\figureheight{0cm}
\caption{Temperature dependence of ESR spectra at 716.7 GHz.
Arrows show T1 and T2 modes.}
\label{fig:3}
\end{figure}

\begin{figure}
\figureheight{0cm}
\caption{The angular dependence of the T1 and T2 modes at 1.6 K in the ${\mib {ac}}$-plane. Numbers for each spectrum is the
angle ${\theta }$ between magnetic field and the ${\mib a}$-axis. The inset shows the angular dependence of zero-field energy gaps of
T1(closed circle) mode and T2(closed rectangles) mode.}
\label{fig:4}
\end{figure}

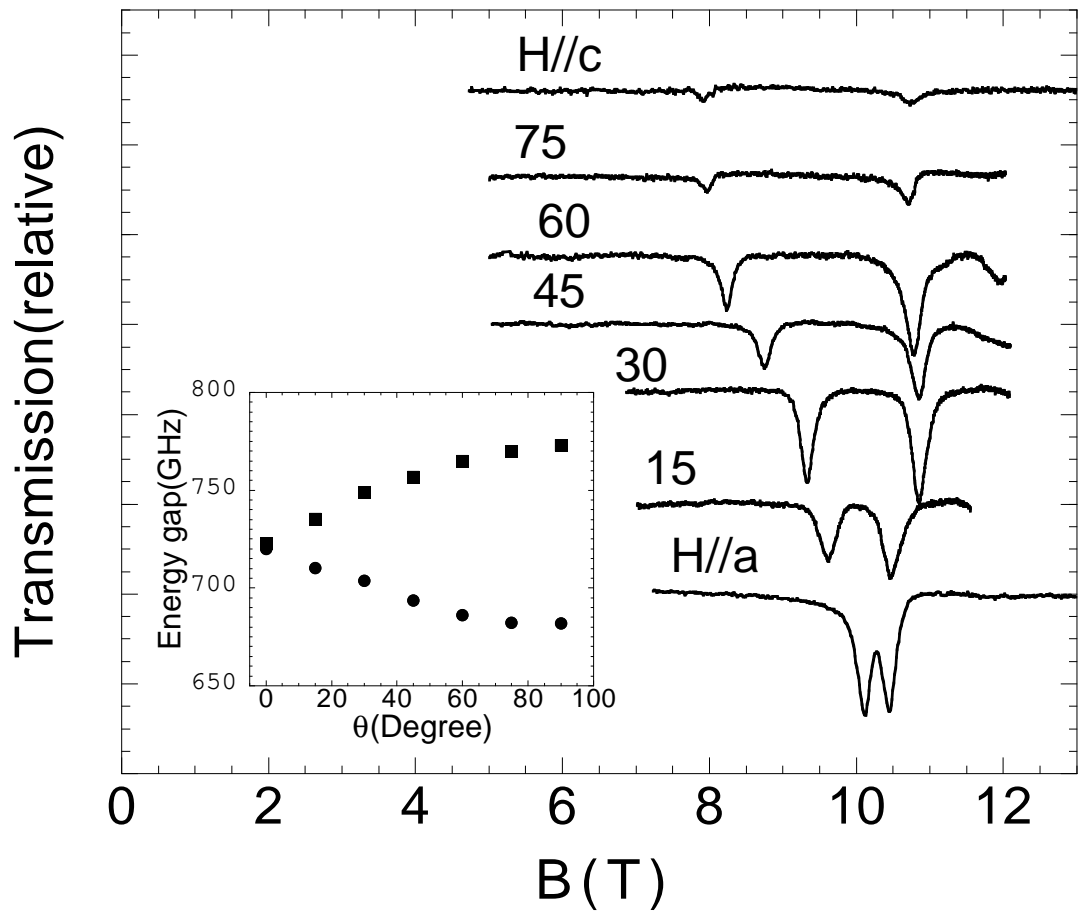
\end{document}